\documentclass{article}

\usepackage[utf8]{inputenc}
\usepackage{amsmath}
\usepackage{authblk}

\title{Circular geodesics stability in a static black hole in new massive gravity}

\author[1]{Andr\'es Ace\~na \thanks{acena.andres@conicet.gov.ar}}
\author[2,3]{Ericson L\'opez}
\author[3]{Franklin Ald\'as}

\affil[1]{Facultad de Ciencias Exactas y Naturales, Universidad Nacional de Cuyo, CONICET, Mendoza, Argentina}
\affil[2]{Observatorio Astronómico de Quito, Unidad de Gravitación y Cosmología, Escuela Politécnica Nacional, Quito, Ecuador}
\affil[3]{Departamento de F\'isica, Facultad de Ciencias, Escuela Politécnica Nacional, Quito, Ecuador}

\date{}

\begin{document}

\maketitle

\begin{abstract}
    We study the existence and stability of circular geodesics in a family of asymptotically AdS static black holes in New Massive Gravity. We show that the existence of such geodesics is determined by the sign of the hair parameter. For a positive hair parameter the stability regions follow the usual pattern, with an innermost unstable null geodesic separated from the horizon, followed by a region of unstable timelike geodesics and then stable timelike geodesics extending in the asymptotic region.
\end{abstract}

\section{Introduction}\label{sec:introduction}

New Massive Gravity (NMG), proposed in 2009 by Bergshoeff, Hohm and Townsend \cite{Berg} (also called BHT massive gravity), has received a great deal of attention due to its remarkable properties, particularly in the context of the AdS/CFT correspondence conjecture and because a variety of exact solutions have been found (see for example \cite{Berg2,Clement,Oliva,Ahmedov2011,Flory2013}). The theory  describes gravity in a vacuum (2+1)-spacetime with a massive graviton. 

In this paper, we focus on the asymptotically AdS rotating black hole solution found in \cite{Oliva}. This solution has a hair parameter and the rotational parameter satisfies $|a| < l$, where the parameter $l$ is related with the cosmological constant as $\Lambda=-l^{-2}$. The extreme rotating case of this NMG black hole can be included after making a change in the hair parameter as suggested in \cite{Giribet}. The extreme case is obtained when $|a|=l$.  We are interested in studying the existence and stability of circular geodesics in the static case of this spacetime. Geodesics in solutions to NMG have been studied profusely, as they provide important information on the properties of the spacetimes. Such studies have been carried out for example for the BTZ black hole \cite{Cruz1994}, for Lifschitz black holes \cite{Cruz2013}, and from the perspective of deviation angles of null geodesics in the same spacetime as this paper focus on \cite{Nakashi2019}. We are also interested in looking for relations between circular geodesics and isoperimetric surfaces, which were studied in \cite{Acena2018}, although the results are discouraging. For analyzing the stability we follow \cite{Cardoso2009}, obtaining the corresponding principal Lyapunov exponent for the geodesics.

The paper is organized as follows. In section \ref{sec:NMG} the family of rotating black holes in NMG is presented. Then conditions for the existence and stability of circular geodesics are discussed in section \ref{sec:geodesics}. We restrict ourselves to the static case in section \ref{sec:static}, finding the timelike and null circular geodesics and determining their stability. There we also consider briefly the non-existent relation with isoperimetric surfaces. Finally, the conclusions are presented in section \ref{sec:conclusions}.

\section{New massive gravity and black hole solutions}\label{sec:NMG}

The action for NMG is
\begin{equation}
S=\frac{1}{16\pi G}\int d^3 x\sqrt{-g}\left(R-2\lambda-\frac{K}{m^2}\right),
\end{equation}
where
\begin{equation}
 K=R_{\mu\nu}R^{\mu\nu}-\frac{3}{8}R^2,    
\end{equation}
and $G$ is the gravitational constant in a (2+1)-spacetime, while $m$ and $\lambda$ are parameters related with the cosmological constant \cite{Berg, Oliva}. In the limit $m^2\rightarrow \infty$ or if the scalar $K$ is equal to zero the action of General Relativity is obtained. As shown in \cite{Oliva}, the field equations in this theory are of fourth order and read
\begin{equation}
G_{\mu\nu}+\lambda g_{\mu\nu}-\dfrac{1}{2m^2}K_{\mu\nu}=0,
\end{equation}
where
\begin{eqnarray}
    K_{\mu\nu} & = & 2\nabla_\rho \nabla^\rho R_{\mu\nu}-\tfrac{1}{2} \left( \nabla_{\mu} \nabla_{\nu} R +g_{\mu\nu}\nabla_\rho \nabla^\rho R\right) \\
	&& -8R_{\mu\rho}R^{\rho}\,_{\nu}+\tfrac{9}{2}R R_{\mu\nu} \\
	&& + g_{\mu\nu}\left(3R^{\rho\lambda}R_{\rho\lambda}-\tfrac{13}{8}R^2\right),
\end{eqnarray}
and $K=g^{\mu\nu}K_{\mu\nu}$.

A plethora of solutions have been found for the field equations, some of them cited in section \ref{sec:introduction}. We concentrate in the asymptotically AdS stationary black hole family obtained in \cite{Oliva}, given in the form presented in \cite{Giribet}, which includes the extreme rotating case. The metric is
\begin{equation}
ds^2=-{N}{F}dt^2+\frac{dr^2}{{F}}+r^2(d\phi+{N}^{\phi}dt)^2
\end{equation}
with
\begin{equation}
{N}=\left[1+\frac{{b}l^2}{4{\sigma}}(1-\xi)\right]^2,
\end{equation}
\begin{equation}
{N}^{\phi}=-\frac{a}{2r^2}(\mu-{b}{\sigma}),
\end{equation}
\begin{equation}
{F}=\frac{{\sigma}^2}{r^2}\left[\frac{{\sigma}^2}{l^2}+\frac{{b}}{2}(1+\xi){\sigma}+\frac{{b}^2 l^2}{16}(1-\xi)^2-\mu\xi\right],
\end{equation}
\begin{equation}
{\sigma}=\left[r^2-\frac{\mu}{2}l^2(1-\xi)-\frac{{b}^2 l^4}{16}(1-\xi)^2\right]^{1/2},
\end{equation}
\begin{equation}
\xi^2=1-\dfrac{a^2}{l^2}.
\end{equation}
Here $\mu=4GM$, being $M$ the mass measured with respect to the zero mass black hole, the angular momentum is given by $J=Ma$,  and ${b}$ is the hair parameter. The rotational parameter $a$ satisfies $-l\leq a \leq l$ and the extreme case is obtained when $|a|=l$. The parameter $l$ is the AdS radius, related with the cosmological constant in the usual way, $\Lambda=-l^{-2}$. The mass parameter $\mu$ is bounded from below,
\begin{eqnarray}
    && \mbox{if }b\leq 0 \Rightarrow \mu \geq -\frac{{b}^2l^2}{4}, \\
    && \mbox{if }b>0 \Rightarrow \mu \geq -\frac{{b}^2l^2}{8}(1-\xi).
\end{eqnarray}
These solutions possess one or more event horizons, an ergosphere, and in general there is a curvature singularity, always hidden by the event horizon. Also, for $b\leq 0$, the extreme limit $|a|=l$ corresponds to a cylindrical end, in all similar to what happens for extreme Kerr. For details of this analysis please refer to \cite{Acena2017} and references therein.

\section{Circular geodesics}\label{sec:geodesics}

If we denote by a dot the derivative with respect to proper time (or the affine parameter for null geodesics) and by $u^\mu$ the tangent vector, the geodesics satisfy
\begin{equation}\label{diffeqn}
-\kappa=g_{\mu\nu}u^{\mu}u^{\nu}= -NF \dot{t}^2 +\frac{\dot{r}^2}{F}+r^2(\dot{\phi}+ N^{\phi}\dot{t})^2
\end{equation}
with
\begin{equation}
\kappa = \begin{cases}
1, \hspace{ 0.5cm}$for timelike geodesics,$\\
0, \hspace{ 0.5cm}$for null geodesics.$
\end{cases}
\end{equation}
The metric possesses the Killing vectors $\partial_t$ and $\partial_\phi$, and the corresponding constants of motion for the geodesics are
\begin{equation}
E = -g_{\mu\nu}\partial_t^{\mu}u^{\nu} = N F \dot{t}-r ^2 N^{\phi}(\dot{\phi} + N^{\phi}\dot{t}),
\end{equation}
\begin{equation}
L = g_{\mu\nu}\partial_\phi^{\mu}u^{\nu} = r^{2} (\dot{\phi} + N^{\phi}\dot{t}).
\end{equation}
Using these in \eqref{diffeqn} and rearranging terms we obtain the one dimensional radial equation of motion
\begin{equation}
    \dot{r}^2=V_r,
\end{equation}
where we define
\begin{equation}
    V_r=\frac{1}{N}(E+N^{\phi}L)^2-F\left(\frac{L^2}{r^2}+\kappa\right).
\end{equation}
The derivative of $V_r$ with respect to $r$, denoted by a prime, is
\begin{eqnarray}
    V_r' & = & \frac{E+N^\phi L}{N^2}\left[ 2 L N N^{\phi'}-(E+N^\phi L)N'\right] \\
    && + \frac{2L^2F}{r^3}-F'\left(\kappa+\frac{L^2}{r^2}\right).
\end{eqnarray}
If we restrict our attention to circular geodesics, these need to satisfy the conditions
\begin{equation}
    V_r = 0,\qquad V_r' = 0.
\end{equation}

To analyze the stability of the circular geodesics we follow the method based on  Lyapunov exponents presented in \cite{Cardoso2009}, and we refer the reader to that work for details. In the case at hand the principal Lyapunov exponent is given by 
\begin{equation}
    \lambda =\sqrt{\frac{V''_r}{2\dot{t}^2}},
\end{equation}
and the unstable orbits are those that have $V''_r>0$. We can associate an instability timescale (or Lyapunov timescale) to the unstable geodesics, given by
\begin{equation}
    T_\lambda = \frac{1}{\lambda},
\end{equation}
which is a measure of how fast the instability would be noticeable. This can be compared with the orbital timescale,
\begin{equation}
    T_\Omega = \frac{2\pi}{\Omega},
\end{equation}
being $\Omega$ the angular velocity of the geodesic
\begin{equation}
\Omega = \frac{\dot\phi}{\dot t},
\end{equation}
in order to calculate the corresponding critical exponent
\begin{equation}
    \gamma = \frac{T_\lambda}{T_\Omega}.
\end{equation}

\section{Static black hole}\label{sec:static}

If we consider the static case, $a=0$, and then
\begin{equation}
    \xi = 1,\,\, \sigma = r,\,\, N = 1,\,\, N^\phi = 0,\,\, F = \frac{r^2}{l^2}+br-\mu.
\end{equation}
The metric takes the simple form
\begin{equation}
ds^2=-F dt^2 +\frac{dr^2}{F}+r^2d\phi^2,
\end{equation}
the horizon is located at
\begin{equation}
    r_+ = \frac{l}{2}\left(-lb+\sqrt{l^2b^2+4\mu}\right),
\end{equation}
and the mass parameter $\mu$ has the following ranges according to the sign of $b$,
\begin{equation}
    b \leq 0 \Rightarrow \mu \geq -\frac{l^2b^2}{4}, \qquad b > 0 \Rightarrow \mu\geq 0.
\end{equation}

Considering the geodesics, the constants of motion are
\begin{equation}
  E = F\dot{t},\qquad L = r^2\dot{\phi}. 
\end{equation}
In the null case the equations of motion can be integrated analytically, which has been done in \cite{Nakashi2019}.

\subsection{Timelike circular geodesics}

Now $\kappa =1$, the potential is
\begin{equation}
V_r = E^2-\left(\frac{r^2}{l^2}+br-\mu\right)\left(\frac{L^2}{r^2} + 1\right),
\end{equation}
and its first two derivatives are
\begin{eqnarray}
    V_r' & = & -\frac{2}{l^2}r-b+\frac{L^2}{r^3}(br-2\mu),\\
    V_r'' & = & -\frac{2}{l^2}-\frac{2L^2}{r^4}(br-3\mu). \label{d2V}
\end{eqnarray}
From the conditions $V_r=0$, $V_r'=0$, we have
\begin{equation}\label{L2}
    E^2 = \frac{2F^2}{br-2\mu},\qquad L^2 = \frac{r^3(2r+l^2b)}{l^2(br-2\mu)}.
\end{equation}
Given that $E$ must be real, we need that
\begin{equation}\label{condE}
    br-2\mu> 0.
\end{equation}
It is necessary to separate the analysis according to the sign of $b$. If $b=0$ then there are no circular geodesics. If $b>0$, the condition form \eqref{condE} is
\begin{equation}\label{eqrE}
    r > r_E = \frac{2\mu}{b}.
\end{equation}
It can be checked that $r_E \geq r_+$. It is also necessary that $L^2>0$, which means
\begin{equation}
    2r+l^2b >0,
\end{equation}
that is
\begin{eqnarray}\label{eqrL}
    r > r_{L} = -\frac{l^2b}{2},
\end{eqnarray}
which is always satisfied, as $r_L<0$. If $b<0$ then condition \eqref{eqrL} stays the same, but condition \eqref{eqrE} is replaced by
\begin{equation}
    r < r_E = \frac{2\mu}{b},
\end{equation}
and it can be checked that
\begin{equation}
    r_E < r_L,
\end{equation}
which implies that there are no circular geodesics. To summarize, if $b\leq 0$ then there are no circular geodesics, if $b>0$ there are circular geodesics for $r>r_E$, and there are no circular geodesics for $r_E>r>r_+$.

Having found the circular geodesics, we turn our attention to the question of stability. From \eqref{d2V} and \eqref{L2} the stability condition, $V_r''<0$, takes the form
\begin{equation}
    3br^2+(l^2b^2-8\mu)r-3l^2b > 0.
\end{equation}
The roots of the quadratic equation are
\begin{equation}
    r_{s\pm} = \frac{1}{6b}\left(-l^2b^2+8\mu\pm\sqrt{l^4b^4+20l^2b^2\mu+64\mu^2}\right),
\end{equation}
and considering that $b>0$ and $\mu\geq 0$ we have
\begin{equation}
    r_{s-}\leq 0,\qquad r_{s+}\geq r_E.
\end{equation}
This means that for $b>0$ there are three regions outside the horizon. For $r_+<r<r_{E}$ there are no circular geodesics, for $r_{E}<r<r_{s+}$ there are circular geodesics but they are unstable, and for $r_{s+}<r$ there are circular geodesics and they are stable.

The angular velocity for a given geodesic is
\begin{equation}\label{angVel}
\Omega=\frac{\dot{\phi}}{\dot{t}} = \frac{LF}{Er^2} = \pm\sqrt{\frac{1}{l^2}+\frac{b}{2r}}.
\end{equation}
It is interesting to note that the angular velocity is a decreasing function of the distance to the horizon, as usually expected, but that it has a non-zero asymptotic value, which is plainly the inverse of the AdS radius $l$.

\subsection{Null circular geodesics}

Considering now the null case, $\kappa = 0$, the position of the null geodesics can be found by the condition $V_r'=0$, which is equivalent to
\begin{equation}
    2F-rF'=0,
\end{equation}
and then there is one null geodesic located at
\begin{equation}\label{nullGeo}
    r_n = \frac{2\mu}{b}.
\end{equation}
If $b=0$ then there is no circular geodesic, if $b<0$ then $r_n\leq r_+$ and there is also no circular geodesic. For $b>0$ the relation between $E$ and $L$ is obtained from $V_r=0$,
\begin{equation}
    \frac{E}{L}=\pm\frac{\sqrt{F}}{r} = \pm\sqrt{\frac{1}{l^2}+\frac{b^2}{4\mu}} = \Omega_n.
\end{equation}
The second derivative of the potential is
\begin{equation}
    V_r'' = \frac{L^2b^4}{8\mu^3} > 0
\end{equation}
and therefore the null geodesic is unstable. It can be seen that the null geodesic is the limit case of the timelike geodesics, being the innermost circular geodesic. The other quantities of interest regarding the instability of the geodesic can be readily calculated,
\begin{equation}\label{critExp}
    \lambda_n = \Omega_n\sqrt{\mu},\qquad \gamma_n = \frac{1}{2\pi\sqrt{\mu}}.
\end{equation}

\subsection{Isoperimetric surfaces}

In this brief section we want to explore if it is possible to compare the ranges and behaviour of circular geodesics with the isoperimetric structure of the spacetime, which was studied in \cite{Acena2018}.

Regarding the isoperimetric structure, all circles of constant radius on a constant-$t$ slice of the spacetime are isoperimetric surfaces. For $b\leq 0$ all of them are also stable. For $b>0$ the isoperimetric surfaces are stable in the range $r_+<r<r_c$, with
\begin{equation}
    r_c = \frac{2}{b}(1+\mu),
\end{equation}
and are unstable for $r>r_c$.

There seems to be no direct relation between the ranges of existence and stability of circular geodesics and the properties of the isoperimetric surfaces.

\section{Conclusions}\label{sec:conclusions}

We have analyzed the existence and stability of timelike and null circular geodesics in a family of asymptotically AdS static black hole solutions of NMG. We conclude that the sign of the hair parameter decides the existence of such geodesics. For $b\leq 0$ there are no circular geodesics. For $b>0$ there is a region close to the horizon without circular geodesics followed by an unstable null geodesic, a region of unstable timelike geodesics and a region of stable timelike geodesics. The $b=0$ case is the well known BTZ black hole, for which the geodesics can be obtained in analytic form \cite{Cruz1994}. Qualitatively the case $b<0$ is closely related with the BTZ black hole, and differs with the $b>0$ case. This was also observed regarding the structure of the horizon with respect to the extremely rotating case in \cite{Acena2017} and regarding the isoperimetric structure of the spacetime in \cite{Acena2018}.

An interesting phenomenon in the $b>0$ case was analyzed in \cite{Nakashi2019}, namely that due to the hair parameter a repulsive behavior of the black hole on null geodesics was observed. This also appears in our analysis, although in a subtle form. For $b\leq 0$ there are no circular geodesics, which can be interpreted as the gravitational attraction to be "too attractive". When the hair parameter is positive, it has a repulsive effect, that allows for the existence of circular geodesics, counteracting the pull towards the black hole. A quantitative measure of this effect can be seen from the position of the null circular geodesic \eqref{nullGeo}, which is closer to the horizon the bigger $b$ gets.

It is also interesting to note that from \eqref{angVel} the velocity profile for test particles orbiting the black hole can be obtained. Inverting the argument, if the velocity profile is known, then the parameters $l$ and $b$ of the spacetime can be obtained. Also, the innermost circular orbit allows us to obtain the value of $\mu$, and therefore from the velocity profile all the parameters of the black hole spacetime can be obtained. Another way of obtaining the value of $\mu$ is through the critical exponent \eqref{critExp}, which surprisingly depends only on $\mu$.

Finally, an attempt was made to relate the geodesics to the isoperimetric surfaces, without success. It is far from clear if there should or should not be any relation between these quantities.

\end{document}